\documentclass[twoside,web]{ieeecolor}
\usepackage{generic}
\usepackage{cite}
\usepackage{amsmath,amssymb,amsfonts}
\usepackage{algorithmic}
\usepackage{graphicx}
\usepackage{textcomp}
\usepackage{amsmath}
\usepackage[margin=0.6in]{geometry}
\usepackage{acronym}
\usepackage{array, makecell}
\newacro{AFM}{antiferromagnetically}
\newacro{LIF}{leaky integrate-and-fire}
\newacro{PMA}{perpendicular magnetic anisotropy}
\newacro{SAF}{synthetic antiferromagnet}
\newacro{FM}{ferromagnet}
\newacro{MTJ}{magnetic tunnel junction}
\newacro{STT}{spin-transfer-torque}
\newacro{DW}{domain wall}
\newacro{HM}{heavy metal}
\newacro{LLGS}{Landau-Lifshitz-Gilbert-Slonczewski}
\newacro{SOT}{spin-orbit torque}
\newacro{STT}{spin-transfer torque}
\newacro{TL}{top layer}
\newacro{BL}{bottom layer}
\newacro{CIP}{current-in-plane}
\newacro{CPP}{current-perpendicular-to-plane}
\newacro{DMI}{Dzyaloshinskii-Moriya interaction}
\newacro{NM}{non-magnetic}
\def\BibTeX{{\rm B\kern-.05em{\sc i\kern-.025em b}\kern-.08em
    T\kern-.1667em\lower.7ex\hbox{E}\kern-.125emX}}
\begin{document}
\title{Self-reset schemes for Magnetic domain wall-based neuron}
\author{Debasis~Das and~Xuanyao Fong, \emph{Member, IEEE}	
\thanks{The authors are with the Dept. of Elect. \& Comp. Engineering, National University of Singapore,
Singapore, 117583 (e-mail: \{eledd,kelvin.xy.fong\}@nus.edu.sg).}
}

\maketitle

\begin{abstract}
Spintronic artificial spiking neurons are promising due to their ability to closely mimic the \acf{LIF} dynamics of the biological LIF spiking neuron.
However, the neuron needs to be reset after firing.
Few of the spintronic neurons that have been proposed in the literature discuss the reset process in detail.
In this article, we discuss the various schemes to achieve this reset in a magnetic \acf{DW} based spintronic neuron in which the position of the \ac{DW} represents the membrane potential.
In all the spintronic neurons studied, the neuron enters a refractory period and is reset when the \ac{DW} reaches a particular position.
We show that the self-reset operation in the neuron devices consumes energy that can vary from of several pJ to a few fJ, which highlights the importance of the reset strategy in improving the energy efficiency of spintronic artificial spiking neurons.

\end{abstract}


\section{Introduction}
\label{sec:introduction}

Proposals for spintronics-based biological spiking \acf{LIF} neuron \cite{li2017magnetic,das2022bilayer} have demonstrated a promising pathway toward neuromorphic computing hardware that emulates the computations of the biological brain. 
In earlier works \cite{sengupta2016magnetic, zhang2016stochastic,srinivasan2016magnetic,liyanagedera2017stochastic,jaiswal2019robustness,grollier2020neuromorphic}, it has been demonstrated that the spintronics devices such as \ac{MTJ}, can mimic the stochastic behavior of the neuron and synapses. 
After firing, the neuron enters a \textit{refractory period} in which it is unresponsive to input spikes and its membrane potential is reset.
The reset mechanism for binary neurons made by \ac{MTJ} has been demonstrated in the existing literature \cite{srinivasan2017magnetic,srinivasan2016magnetic}.
However, apart from such binary operations, magnetic \ac{DW} based can also be used to mimic the neuronal \cite{hassan2018magnetic,sengupta2016proposal} as well as synaptic \cite{zand2018low,sengupta2017encoding,sengupta2018neuromorphic} operations.
With the increasing demand for energy-efficient neuromorphic computing hardware, there is hence an urgent need to address the issue of resetting the membrane potential in spintronic neuronal devices, which, as we will show in this work, can consume significant amounts of energy if not carefully designed.

In this article, we consider magnetic \ac{DW} based \ac{LIF} neurons where the membrane potential corresponds to the position of a magnetic domain in the device. 
Once the magnetic domain reaches the detector, it activates a certain mechanism based on the reset scheme, either by nucleating a new domain at the generator region or by sending a reset signal that sends the magnetic domain to its initial position. 
Based on the auto-activation of the reset mechanism upon firing, we propose various schemes to design a self-reset mechanism using monolayer as well as \ac{SAF} bilayer devices.  
The neurons studied in this work are implemented using monolayer \acf{FM} \cite{brigner2019graded,mah2021domain} and \acf{AFM} coupled bilayer system \cite{das2022bilayer,zhang2016magnetic}.
Micromagnetic simulations were used to study the dynamics of the magnetic domain that acts as the membrane potential of the \ac{LIF} neuron.
The total energy required to complete the entire rest-to-fire-to-reset process is calculated so as to compare various self-reset schemes for the spintronic-based \ac{LIF} neuron. 

The rest of the article is organized as follows. In Section~\ref{sec:Device_model_math}, we describe various devices that we use to implement the working principle of an artificial neuron along with the mathematical framework used in our simulation. 
Detailed device operations and corresponding results for all the proposed devices are discussed in Section~\ref{sec:results}, and Section~\ref{sec:conclusion} concludes the article.  

\section{Device concepts \& Model} \label{sec:Device_model_math}

In this section, we first describe the device structures used to implement the functionality of a \ac{LIF} neuron in Section~\ref{sec:device}. 
Thereafter, in Section~\ref{sec:model} we describe the micromagnetic model for simulating the magnetization dynamics of the device. 

\begin{figure}[!b]
	\centerline{\includegraphics[scale=0.6]{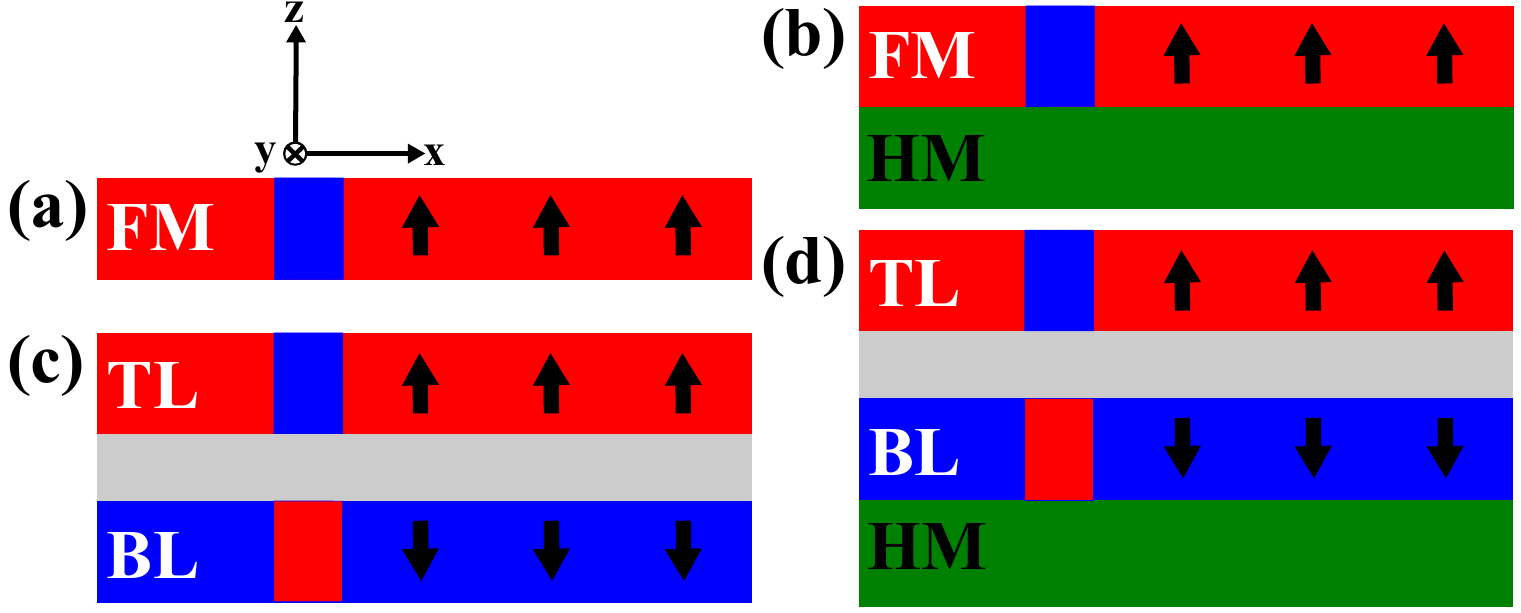}}
	\caption{Side-view of the schematic of (a)-(b) monolayer device; (c)-(d) bilayer device, along with the magnetic domain. Upper (lower) arrows in show magnetization direction along +(-)z-direction. }
	\label{Fig_1:devices}	
\end{figure}

\subsection{Device structure}\label{sec:device}

Structures such as monolayer and \ac{AFM} coupled bilayer structures (shown in Fig.~\ref{Fig_1:devices}) were considered for possible implementations of self-resetting spintronic \ac{LIF} neuron. 
The monolayer device consists of a single \ac{FM} layer as shown in Fig. \ref{Fig_1:devices}(a) and (b), whereas in a \ac{SAF} bilayer system, two \ac{FM} layers---\acf{TL} and \acf{BL}---are separated by a thin \ac{NM} layer as shown in Fig. \ref{Fig_1:devices} (c) and (d). 
All the \ac{FM} layers are assumed to have \acf{PMA}.
In a bilayer system, the magnetization of \ac{TL} and \ac{BL} are coupled through an \ac{AFM} exchange interaction and due to \ac{PMA}, the magnetization of these two layers are oriented along the $+z$ and $-z$ direction, respectively.

The position of the magnetic domain represents the membrane potential, and it can be driven either by \acf{STT} or \acf{SOT}.
\ac{STT} is generated by injecting a spin-polarized current into \ac{FM} in Fig.~\ref{Fig_1:devices}(a) and \ac{BL} in Fig.~\ref{Fig_1:devices}(c), which induces a Zhang-Li torque \cite{zhang2004roles} that moves the \ac{DW} opposite to the direction of the current.
Alternatively, injection of a charge current (along the +$x$-direction) into the \acf{HM} layer beneath the \ac{FM} layer in Fig.~\ref{Fig_1:devices}(b) and \ac{BL} in Fig.~\ref{Fig_1:devices}(d) gives rise to a $+z$-directed spin current (polarized along the $y$-direction), which is injected into the adjacent \ac{FM}/\ac{BL} layer \cite{sengupta2016proposal} that moves the \ac{DW} using \acf{SOT} . 

\subsection{Micromagnetic model}\label{sec:model}

The Hamiltonian for a magnetic layer is given as
\begin{equation}
	\begin{split}
		H^l=&K_u^l\sum_i\left(1-\left(\mathbf{m}_{i}^{l}\cdot\mathbf{z}\right)^2\right) + H_\text{dipole}^{l}\\ &
		+ A_\text{intra}\sum_{\langle{}i,j\rangle}\mathbf{m}_i^l \cdot \frac{\mathbf{m}_i^l - \mathbf{m}_j^l}{\Delta_{ij}}
		\label{Eq:Hamiltonian}
	\end{split}
\end{equation}  
$l$ represents the index of the magnetic layer in the device.
$\mathbf{m}_{i}^{l}$ represents the local normalized magnetization vector, and $\langle{}i,j\rangle$ denotes the nearest neighbour interaction in layer $l$.
The first term in Eq.~(\ref{Eq:Hamiltonian}) represents the energy for \ac{PMA}, where $K_u^l$ is the anisotropy constant for layer $l$. 
The second term represents the energy due to dipole-dipole interactions.
The third term represents the intra-layer exchange interaction, where $A_\text{intra}$ is the exchange stiffness constant and $\Delta_{ij}$ is the discretization size between cell $i$ and cell $j$.
In the devices shown in Fig.~\ref{Fig_1:devices} (b) and (d), the \ac{HM} attached to the magnetic layer may induce \acf{DMI} with energy given by 
\begin{equation}
	H_{DMI}=\sum_{\langle{}i,j\rangle}\mathbf{D}\cdot\left(\mathbf{m}_i^l \times \mathbf{m}_j^l\right)
\end{equation}
where $\mathbf{D}$ is the DM vector whose magnitude is denoted by $D$. 
In the bilayer system, there is also an additional energy due to inter-layer \ac{AFM} coupled exchange interaction between \ac{BL} and \ac{TL}, which is given by
\begin{equation}
H_\text{inter}=\sigma \sum_k \frac{1-\mathbf{m}_k^{TL} \cdot \mathbf{m}_k^{BL}}{\Delta_k}
\label{Eq:Exchange_energy}
\end{equation}
where $\sigma$ is the bilinear surface exchange coefficients between the two layers, and $\Delta_{k}$ is the discretization cell size in the direction from cell $k$ of \ac{BL} towards cell $k$ of \ac{TL}.
Thus, the net Hamiltonian for the different systems considered in this work is given by
\begin{equation}
	H^{net}=\begin{cases} H^{FM} & \text{monolayer }, \\
	H^{FM}+ H_{DMI} & \text{monolayer+\ac{HM}},\\
	H^{TL}+H^{BL}+H_{inter} & \text{bilayer},\\
	H^{TL}+H^{BL}+H_{inter}+H_{DMI} & \text{bilayer+HM},	
	\end{cases}
	\label{Eq:H_net}
\end{equation} 
which provides an effective magnetic field $\mathbf{H}_\text{eff}^{l}=-\partial{}H^{net}/\partial{}\mathbf{m}^l$ acting on layer $l$ of the system.

The magnetization dynamics of the \ac{DW} in a layer $l$ is simulated by solving the \acf{LLGS} equation under the influence of $\mathbf{H}_\text{eff}^{l}$, given by 
\begin{equation}
	\frac{d\mathbf{m}^l}{dt}=-\gamma \left(\mathbf{m}^l\times \mathbf{H}_\text{eff}^l\right)+\alpha \left(\mathbf{m}^l\times\frac{d\mathbf{m}^l}{dt}\right) + \boldsymbol{\tau}^l
	\label{Eq:LLGS}
\end{equation}
where $\gamma=2.211\times10^5~\text{m/(A.s)}$ is the gyromagnetic ratio, and $\alpha$ is the Gilbert damping factor. 
The first and the second terms in Eq. (\ref{Eq:LLGS}) are the precession and damping terms, respectively. $\boldsymbol{\tau}$ is the spin torque acting on layer $l$, which is Zhang-Li torque for the devices without the \ac{HM}, and \ac{SOT} for devices with \ac{HM}. 
Zhang-Li torque is given by 
\begin{equation}
	\boldsymbol{\tau}_{ZL}=-\left(\mathbf{u}\cdot \boldsymbol{\nabla}\right)\mathbf{m}^l + \beta_{STT} \mathbf{m}^l\times\left[\left(\mathbf{u}\cdot \boldsymbol{\nabla}\right)\mathbf{m}^l\right]
\end{equation}
where $\mathbf{u}=\frac{JPg\mu_B}{2eM_S}\hat{i}$ is the velocity vector along the direction of electron motion. Here, $J$ is the current density, $P$ is the spin polarization, $g$ is the Land\'e factor, $\mu_B$ is Bohr magneton, $e$ is the electronic charge, $M_S$ is the saturation magnetization, and $\hat{i}$ is the unit vector along $x$-direction denoting the direction of the electron flow.
$\beta_{STT}$ is a factor related to non-adiabatic \acf{STT}.

Alternatively, \ac{SOT} is given by 
\begin{equation}
	\boldsymbol{\tau}_{SOT}=-\gamma \beta_{SOT} \left[ \epsilon \left(\mathbf{m}^l\times\mathbf{m}^l\times \mathbf{m}_p\right) + \epsilon' \left(\mathbf{m}^l\times \mathbf{m}_p\right)\right]
\end{equation}
where $\beta_{SOT}=\hbar J_{HM}/\left(\mu_0 e t_z M_S \right)$, $\hbar$ is reduced Planck's constant, $J_{HM}$ is the charge current density injected to the \ac{HM}, $\mu_0$ is free space permeability, $e$ is the electronic charge, and $t_z$ is the thickness of the magnetic layer above the \ac{HM}. $\epsilon=P\Lambda^2/\left(\left(\Lambda^2+1\right)+\left(\Lambda^2-1\right)\left(\mathbf{m}^l\cdot\mathbf{m}_p\right)\right)$, where $\Lambda\geq1$, and $\mathbf{m}_p$ is the direction of the spin polarization. $\epsilon'$ is the secondary spin transfer term.

\section{Result \& Discussion}\label{sec:results}

The Object Oriented Micromagnetic Framework (OOMMF) was used to perform micromagnetic simulations of all devices in this work. 
The dimensions of all magnetic layers (FM, TL, BL) are 512~nm$\times$32~nm$\times$1~nm discretized into 1~nm side cubes. 
Parameters used in the simulations are listed in Table \ref{Table:parameters}.
\begin{table}[t]
	\centering
	\caption{\label{tab:table1}%
		Device simulation parameters
	}
	\label{Table:parameters}
	\begin{tabular}{|c|c|}	
		\hline		
		\textrm{Parameters} & \textrm{Value}\\
		\hline
		Saturation magnetization, $M_s$ & 580 kA/m  \\ 
		Gilbert damping factor, $\alpha$ & 0.3 \\
		Spin polarization factor, $P$ & 0.4 \\
		PMA constant, $K_u$ & 0.8 $\mathrm{MJ/m^3}$ \\
		Anisotropy gradient, $\Delta K_u$ & 0.78125 $\mathrm{GJ/m^4}$\\
		Exchange stiffness constant, $A_\text{intra}$ & 15 pJ/m \\
		\hline
	\end{tabular}	
\end{table}
\begin{figure}[t!]
	\centerline{\includegraphics[scale=0.37]{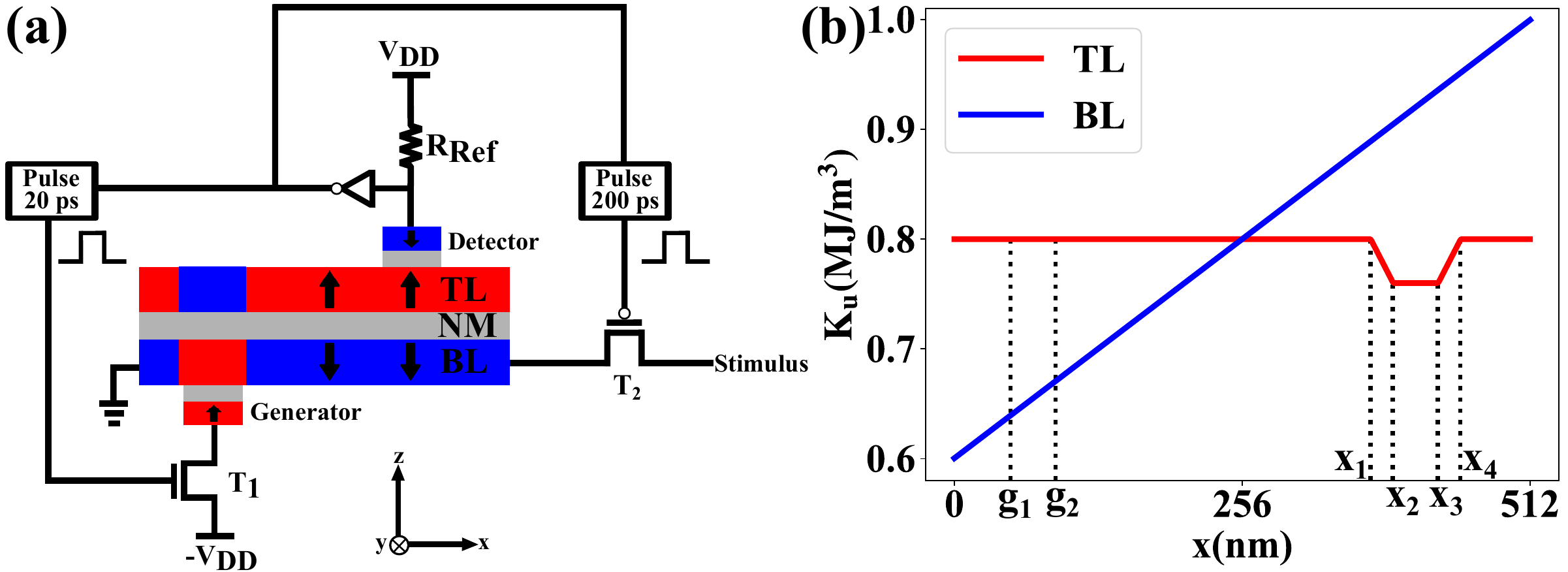}}
	\caption{(a) Schematic of the side view of the device along with the electrical connections from the detector to the generator. (b) The $K_u$ profile along the length of the device. $K_u$ for TL at $\mathrm{x_1}$=370~nm and $\mathrm{x_4}$=450~nm is 0.8~$\mathrm{MJ/m^3}$, whereas it is 0.76~$\mathrm{MJ/m^3}$ at $\mathrm{x_2}$=390~nm and $\mathrm{x_3}$=430~nm. The detector spans from $\mathrm{x_2}$ to $\mathrm{x_3}$ whereas the generator spans from $\mathrm{g_1}$=50~nm to $\mathrm{g_1}$=90~nm.}
	\label{Fig_2:bilayer_self_reset_device}	
\end{figure}

The self-reset mechanism may be achieved using two methods: (i) by nucleating a second domain at the initial position without a reset current, or (ii) by injecting a reset current opposite to the stimulus current to return the magnetic domain to its initial position.
One way to implement a device that achieves self-reset without a reset current is shown in Fig.~\ref{Fig_2:bilayer_self_reset_device}(a).
The device consists of a \ac{SAF} formed by two \ac{AFM}-coupled \acp{FM} layers that are separated by a thin \ac{NM} layer.
The \ac{MTJ}-like structures on the left side of the BL and right side of the TL form the \textit{domain generator} and \textit{detector}, respectively.
The fixed layers of these \acp{MTJ} also possess \ac{PMA}, whereas BL and TL act as the free layers of the corresponding \acp{MTJ}.
A magnetic domain may be generated in \ac{BL} by passing a current through the generator, which is controlled by the transistor, $\mathrm{T_1}$. 
\begin{figure}[t!]
	\centerline{\includegraphics[scale=0.42]{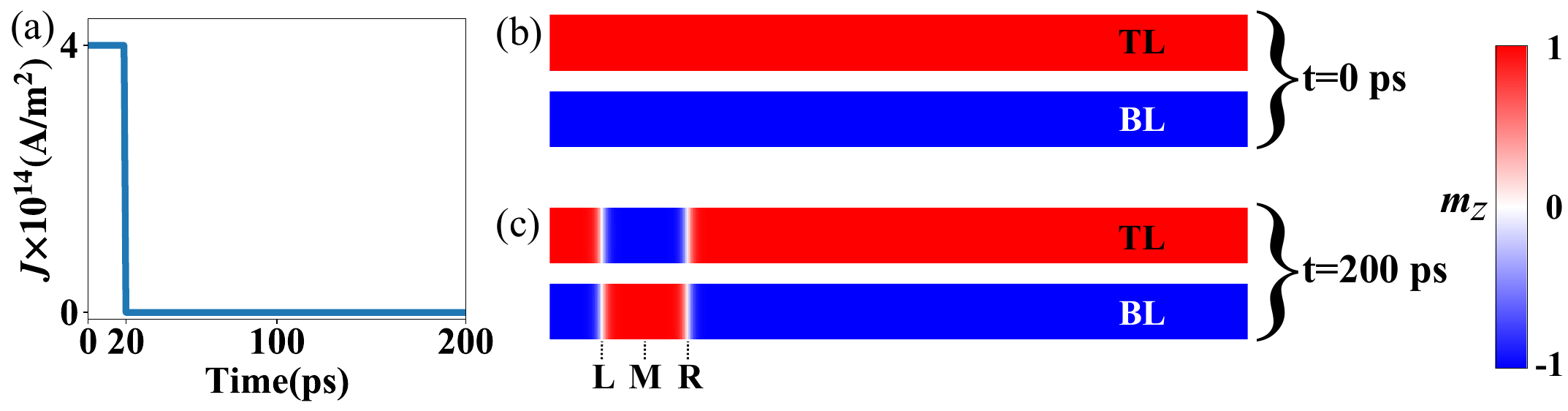}}
	\caption{(a) A 20 ps pulse is applied at the generator region to create the domain. (b)  Two dimensional color plot for $m_z$ on the $x\mbox{-}y$ plane for TL and BL at (b) 0 ps; and (c) 200 ps. }
	\label{Fig_3:bilayer_nucleation_pulse}	
\end{figure}
The widths of the detector and generator are identical to the width of the nanotrack (\emph{i.e.}, 32~nm).
To emulate the leak functionality, the uniaxial anisotropy constant ($K_u$) of BL is designed to be described by \cite{das2022bilayer, das2019skyrmion, li2017magnetic}
\begin{equation}
	K_{u}(x)=K_{u}^{c}+\Delta K_{u}x.
	\label{Eq:Ku_grad}
\end{equation}
This type of anisotropy gradient can be made by ion bombardment \cite{matczak2014tailoring} during the fabrication of the device.  
The $K_u$ profiles of BL and TL along the length of the device are shown in Fig. \ref{Fig_2:bilayer_self_reset_device}(b). Marks on the graph demarcate the locations of the generator and detector.

The neuron needs to be initialized first by injecting +$z$-directed spin-polarized current pulse locally in \ac{BL} through the generator as shown in Fig.~\ref{Fig_3:bilayer_nucleation_pulse}(a) to nucleate the magnetic domains.
This was achieved by applying a 20~ps current pulse of magnitude 4$\times10^{14}~\mathrm{A/m^2}$, which flips the spin orientation of the local spins of the BL to +$z$-direction.
Since BL and TL are AFM-coupled, a -$z$-directed magnetic domain is created in the TL adjacent to the -$z$-directed magnetic domain in the BL.
From Fig.~\ref{Fig_3:bilayer_nucleation_pulse}, we observe that the nucleated magnetic domain is stable even after the removal of the current pulse.
Two-dimensional color plots of $m_z$ on the $x\mbox{-}y$ plane in TL and BL are shown in Fig.~\ref{Fig_3:bilayer_nucleation_pulse}(b) and (c) for the time instances \emph{t}=0 ps (at the beginning of the pulse) and \emph{t}=200 ps, respectively. 
The calculation of the consumed energy can be obtained by the following \cite{bhattacharya2019low,das2022bilayer}
\begin{equation}
    E=\rho L A J^2 t_\text{delay}
\end{equation}
where $\rho$ is the resistivity of the material, $L$ is the length of the device, $A$ is the cross-sectional area through which the current flows, $J$ is the current density, and $t_\textbf{delay}$ is the time required to complete the operation.
The energy consumed to nucleate the domains is approximately 1.11~pJ. 
In Fig.~\ref{Fig_3:bilayer_nucleation_pulse}(c), M, L, and R represent the position of the mid-point, left, and right walls of the magnetic domain, respectively.  
\begin{figure}[!t]
	\centerline{\includegraphics[scale=0.42]{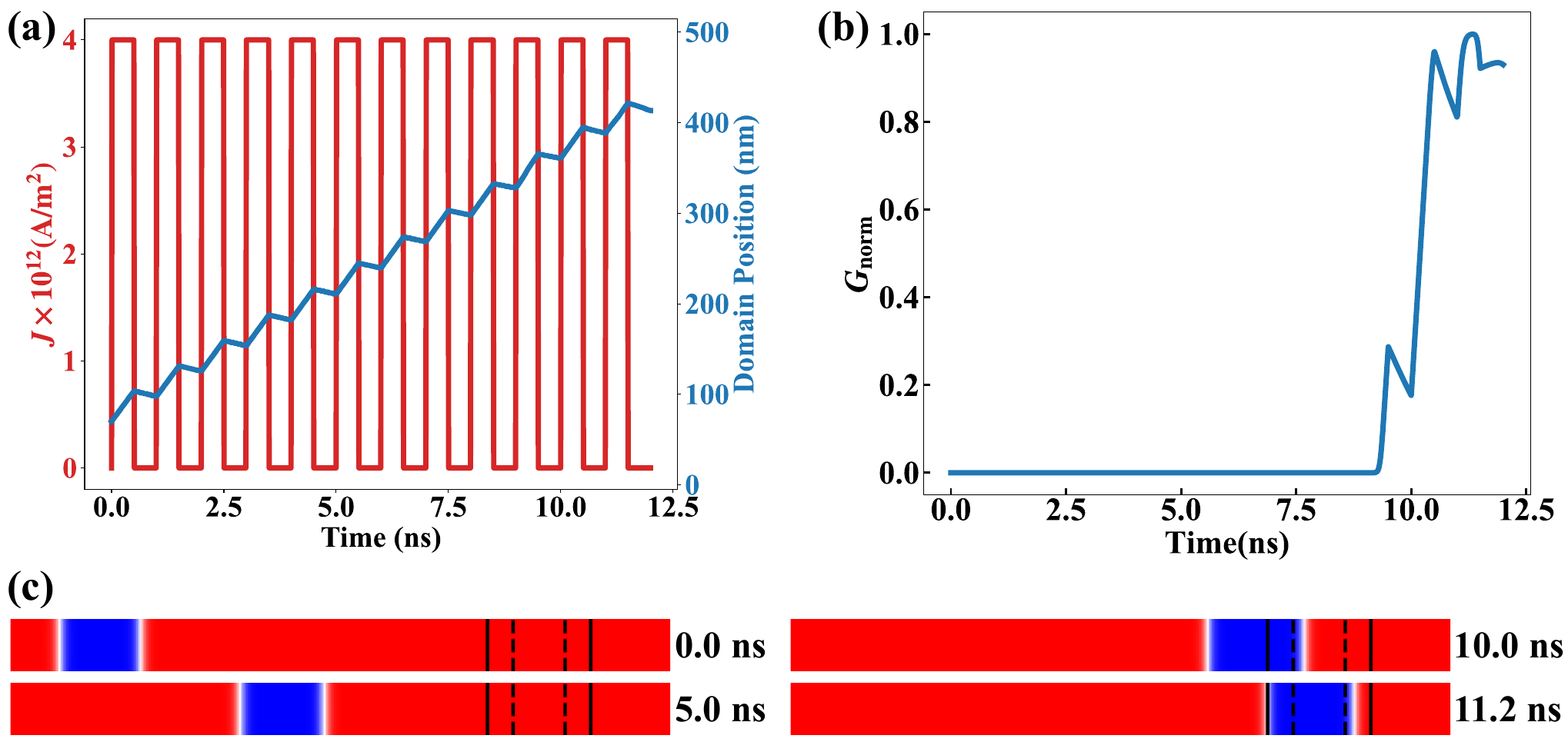}}
	\caption{(a) Plot of current pulse magnitude (red) with time, applied to BL for the domain propagation. Variation of domain mid-position with time under the influence of the applied pulse, imitating the LIF functionality. (b) Variation of normalized conductance of the detector region with time. (c) Snapshot of $m_z$ on the $x\mbox{-}y$ plane for TL at different time instants.}
	\label{Fig_4}	
\end{figure}

The entire rest-to-fire-to-reset operation of the self-resetting neuron can be described in three steps.
In the first step, the input stimulus is injected along the -$x$-direction to the \ac{BL} in the form of a square current pulse train with 0.5~ns pulse width and 1~ns period as shown by the red curve in Fig.~\ref{Fig_4}(a).
In the rest of the paper, we assume the same pulse width and time period for the square current pulse train, applied to the various proposed neuron devices. 
Here, we assume the magnitude of the applied $J$ is $4\times10^{12}~\mathrm{A/m^2}$.
The induced \ac{STT} during the high state of the current pulse pushes the \acp{DW} along +$x$-direction~\cite{zhang2004roles,thiaville2005micromagnetic,parkin2008magnetic}, whereas during the low state of the current pulse, the \acp{DW} move in the opposite direction due to the anisotropy gradient \cite{brigner2019graded} in \ac{BL}. 
This is shown by the corresponding blue graph of the position of M versus time in Fig.~\ref{Fig_4}(a), which confirms that the device performs the leaky and integration functionality of a biological neuron.
After sufficient current spikes, the magnetic domain enters the detector region and becomes pinned due to the $K_{u}$ well in \ac{TL} (refer to Fig. \ref{Fig_2:bilayer_self_reset_device}(b)).
The energy consumed for the continuous spike integration in Fig.~\ref{Fig_4}(a) is approximately 0.4~pJ.
As shown in Fig. \ref{Fig_2:bilayer_self_reset_device}(a), the magnetization of the fixed layer of the detector is pinned in the -$z$-direction.
When the magnetic domain in \ac{TL} reaches under the detector, the detector \ac{MTJ} becomes the parallel state and causes the conductance across it to increase.
The conductance, \textit{G}, of this \ac{MTJ} is calculated using 
\begin{equation}
	G=G_0\sum_i\frac{1+P^2cos\theta_{i}}{1+P^2}
\end{equation}
where \textit{G}$_{0}$ is the conductance when all the spins of the free and fixed layer of the detector unit are perfectly parallel to each other, \textit{P} is the spin polarization, and $\theta_{i}$ is the angle between the magnetization of the \textit{i}-th cell of
the fixed layer of the detector and the corresponding region of TL.
Fig.~\ref{Fig_4}(b) graphs the change of normalized conductance, $\textit{G}_{\mathrm{norm}}=\left(\textit{G}-\textit{G}_{\mathrm{min}}\right)/\left(\textit{G}_{\mathrm{max}}-\textit{G}_{\mathrm{min}}\right)$, with time as the magnetic domain moves.
When the domain is not in the detector region, \textit{G}$_{\mathrm{norm}}$=0.
As the domain enters the detector region, \textit{G}$_{\mathrm{norm}}$ increases and reaches a maximum value at 11.2~ns.
A few snapshots of the $m_{z}$ on the $x\mbox{-}y$ plane for TL at different time instants are shown in Fig.~\ref{Fig_4}(c).
The black solid (dashed) vertical lines are drawn at $x$=370~nm and $x$=450~nm ($x$=390~nm and $x$=430~nm), respectively, to mark the locations where $K_{u}$ is changing in the TL. 
Similar magnetic domains but with opposite polarization are found in the BL (not shown in the figure).
\begin{figure}[!b]
	\centerline{\includegraphics[scale=0.45]{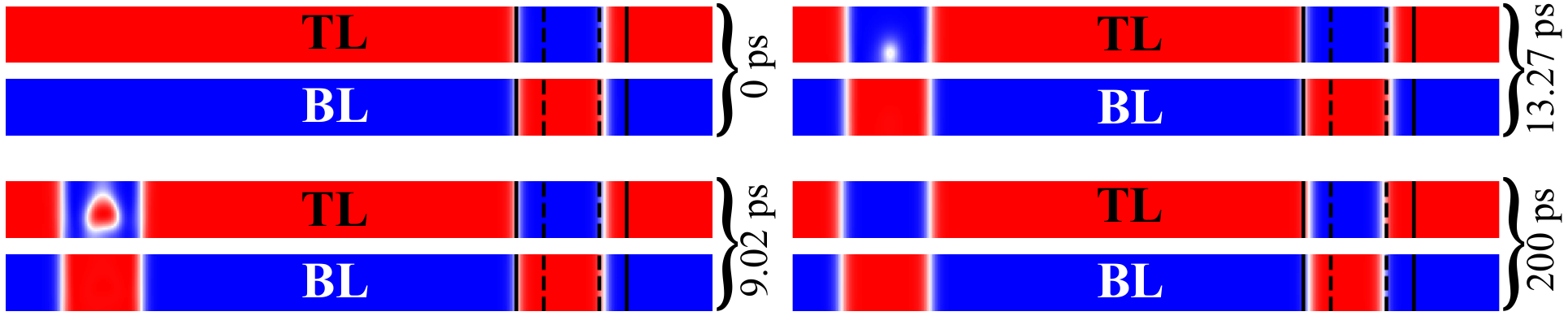}}
	\caption{Snapshot of the $m_z$ on the $x\mbox{-}y$ plane for TL and BL respectively at different time instants.}
	\label{Fig_5}	
\end{figure}

When the magnetic domain is pinned in the detector, the neuron fires and a two-step reset process is triggered by activating two voltage pulses for 20~ps and 200~ps respectively, as shown in Fig. \ref{Fig_2:bilayer_self_reset_device}(a).
One purpose of these pulses are to prevent input stimulus from being injected into \ac{BL} and thereby achieve the refractory behaviour of the LIF neuron.
The other purpose is to perform the first step in the reset process by first nucleating a new magnetic domain in the generator.
The magnetic domain in the detector region is eliminated during the second step of the reset process, which we discuss later.

The 20~ps pulse turns on the transistor $\mathrm{T}_{1}$, causing spin-polarized current to flow through the generator.
Alternatively, the 200~ps pulse deactivates the transistor $\mathrm{T_2}$, ensuring no input stimulus is applied to the \ac{BL} and the neuron enters into the refractory period during this time.
The spin-polarized current through $\mathrm{T_1}$ nucleates a pair of magnetic domains at TL and BL in the generator region just like in the initialization process described earlier.
This process is depicted through the snapshots of $m_{z}$ on the $x\mbox{-}y$ plane at different time instants shown in Fig.~\ref{Fig_5}.
At \textit{t}=0~ps, there are only magnetic domains in TL and BL under the detector region.
As soon as the spin-polarized current enters the BL through the generator, local spins in that region of the BL start to flip---the corresponding spins in TL also flip due to the AFM exchange coupling.
After some time, magnetic domains can be found in the generator and detector regions in the TL and BL, as shown in Fig.~\ref{Fig_5} for \textit{t}=200~ps.

In the final step of the two-step reset process, a single current pulse like the one in Fig.~\ref{Fig_4}(a) is applied to \ac{BL} to eliminate the magnetic domain in the detector.
This current exerts torque on both domains in the detector and the generator regions.
Due to the finite gradient of $K_{u}$ from $x$=430~nm to $x$=450~nm in TL, the right-most \ac{DW} of TL in the detector region encounters an energy barrier that exerts a -$x$-directed force on the \ac{DW} and prevents it from exiting the detector region towards the right.
At the same time, the left-most \ac{DW} of TL in the detector region experiences no such force and is pushed to the right.
Over time, magnetic domains in TL and BL under the detector are squeezed until they are annihilated completely.
\begin{figure}[!t]
	\centerline{\includegraphics[scale=0.4]{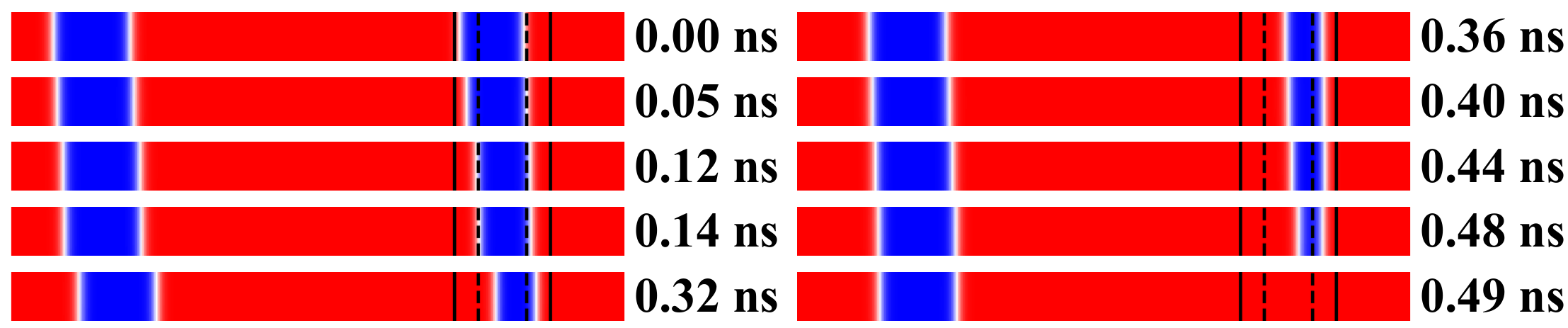}}
	\caption{Snapshot of the $m_z$ on the $x\mbox{-}y$ plane for TL at different time instants.}
	\label{Fig_6}	
\end{figure}
The process of annihilating the magnetic domains in the detector is depicted in Fig.~\ref{Fig_6} through the snapshots of the $m_{z}$ on the $x\mbox{-}y$ plane at different time instants.
Thus, our simulation results confirm that the SAF-based LIF neuron device has the leaky integrate-and-fire behavior and with the desired self-reset behavior during the refractory period.
For the entire rest-to-fire-to-reset operation, the neuron consumes $\sim$1.5~pJ of energy.

\begin{figure}[!b]
	\centerline{\includegraphics[scale=0.4]{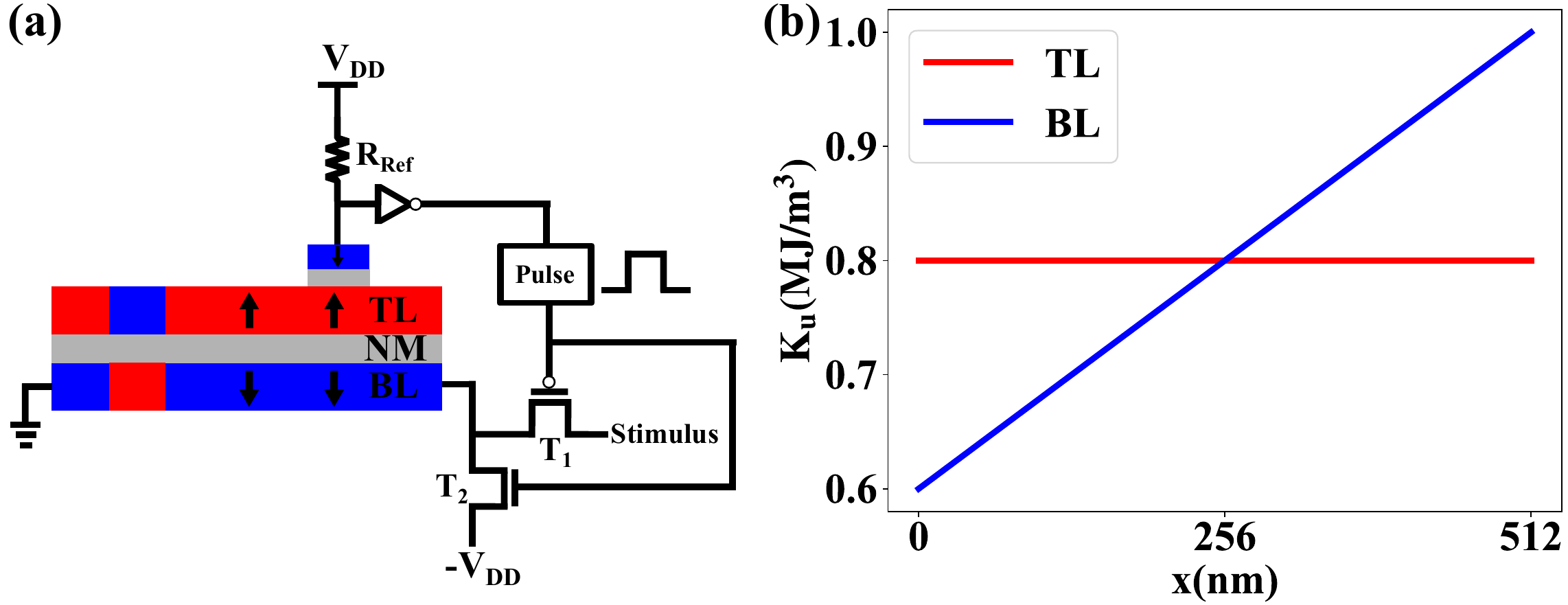}}
	\caption{(a) Schematics of the bilayer device along with reset circuitry. (b) $K_u$ profile along the length of the device.}
	\label{Fig:Bilayer_device_wout_notch}	
\end{figure}

Alternatively, the reset operation in the neuron may be achieved by moving the magnetic domain from the detector back to the generator.
This is achieved by applying a reset current opposite to the stimulus after the neuron fires.
A neuron with this method of self-reset may be implemented using an \ac{SAF} bilayer device without a $K_u$ well in \ac{TL}.
A schematic of the neuron device with the reset circuitry is shown in Fig.~\ref{Fig:Bilayer_device_wout_notch}(a).
An anisotropy gradient in the \ac{BL} is used to implement the leaky behavior in the neuron.
Unlike the previous bilayer device, there is no pinning site in \ac{TL} (see Fig.~\ref{Fig:Bilayer_device_wout_notch}).
As an initial condition, consider a similar magnetic domain position as shown in Fig.~\ref{Fig_3:bilayer_nucleation_pulse}(c). 
The initial magnetic domain can be nucleated by attaching a generator similar to the previous device, which we do not include in the schematic for convenience. 
The position of this magnetic domain represents the membrane potential of the neuron.

\begin{figure}[!t]
	\centerline{\includegraphics[scale=0.19]{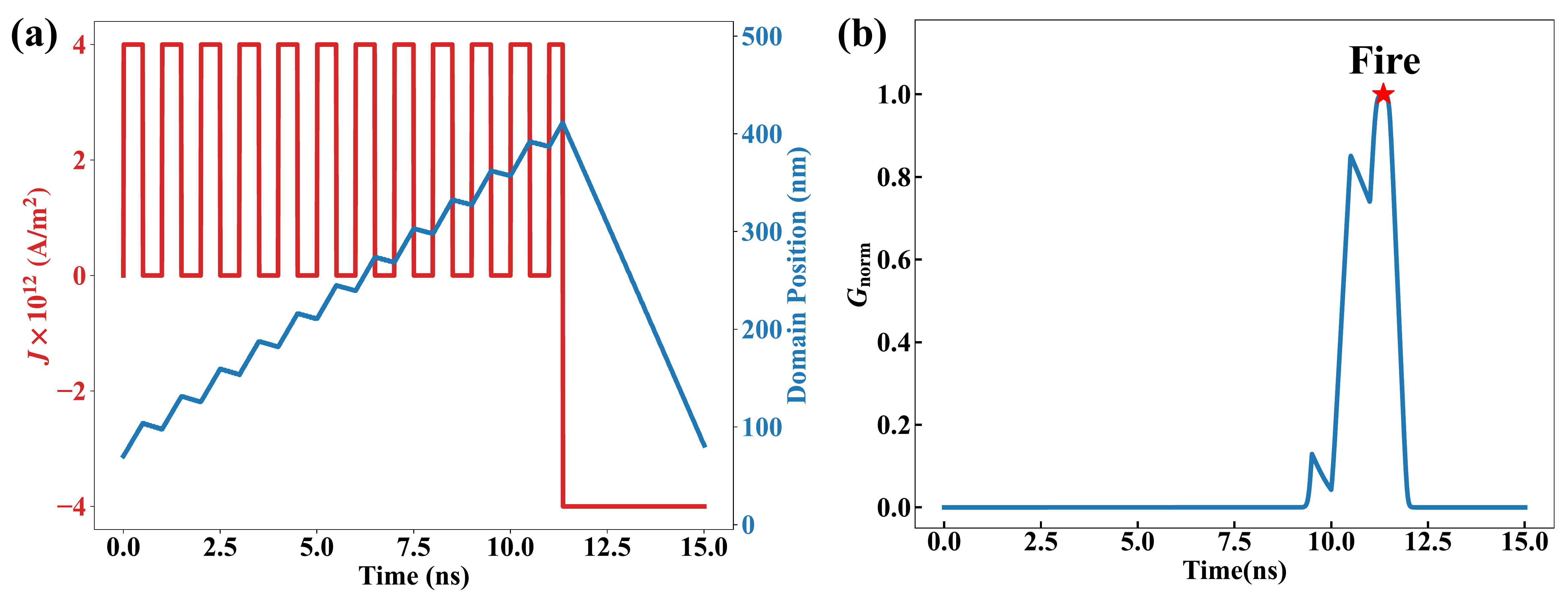}}
	\caption{(a) Plot of current pulse train as the input stimulus injected to the device, along with the reset current after the neuron fires, (b) Variation of the normalized conductance of the detector region with time.}
	\label{Fig_8}	
\end{figure}

To validate the integration functionality, the square pulse train in Fig.~\ref{Fig_8}(a) (red curve) is applied to \ac{BL} via the transistor $\mathrm{T_1}$.
An \ac{MTJ} like detector is placed at the right side of the \ac{TL}, which is connected to a voltage divider circuit via resistance $\mathrm{R_{Ref}}$. 
The magnetic domain moves in the +$x$-direction whenever the magnitude of current is non-zero.
When the current is zero, the magnetic domain moves in the -$x$-direction due to $K_u$ gradient in \ac{BL}, which emulates the leak functionality.
The variation of the mid position of the magnetic domain, M, versus time is shown as the blue plot in Fig.~\ref{Fig_8}(a), which confirms the integrate and leak functionality of the neuron device.
The magnetic domain takes $\sim$11.35~ns to reach the detector, which requires an energy of 378~fJ.
Fig.~\ref{Fig_8}(b) plots the normalized conductance of the detector region, $G_{\mathrm{norm}}$, versus time.
When the domain is outside of the detector region, $G_{\mathrm{norm}}$ is low. 
As the domain enters the detector region, $G_{\mathrm{norm}}$ starts to rise and reaches a maximum at 11.35~ns.
Thereafter, the neuron fires as shown by the red asterisk in Fig.~\ref{Fig_8}(b).

When the neuron fires, the voltage divider circuit triggers a pulse generator circuit to generate a 3.66~ns voltage pulse, which deactivates the transistor $\mathrm{T_1}$ shown in Fig.~\ref{Fig:Bilayer_device_wout_notch}(a).
This blocks any input stimulus, which means the neuron enters the refractory period. 
The pulse is also given to the transistor $\mathrm{T_2}$ shown in Fig.~\ref{Fig:Bilayer_device_wout_notch}(a), which allows a reset current of magnitude $4\times10^{12}~\mathrm{A/m^2}$ to flow along the +$x$-direction to move the domain back to its initial position and thus, resetting the neuron.
The energy consumed by the reset current is 260~fJ.
Thus, for the entire rest-to-fire-to-reset operation, the neuron consumes $\sim$638~fJ of energy.

The neuron devices presented thus far utilize \ac{STT} for domain wall motion and requires pJ level of energy consumption for the entire rest-to-fire-to-reset operation.
The energy consumption may be further reduced if the charge current required is reduced.
This may be achieved using \ac{SOT} generated by a \ac{HM} layer that is adjacent to the \ac{FM} layer \cite{das2022bilayer}. 
Let us now consider neuron devices that utilize \ac{SOT} for magnetic domain motion.
\begin{figure}[!t]
	\centerline{\includegraphics[scale=0.2]{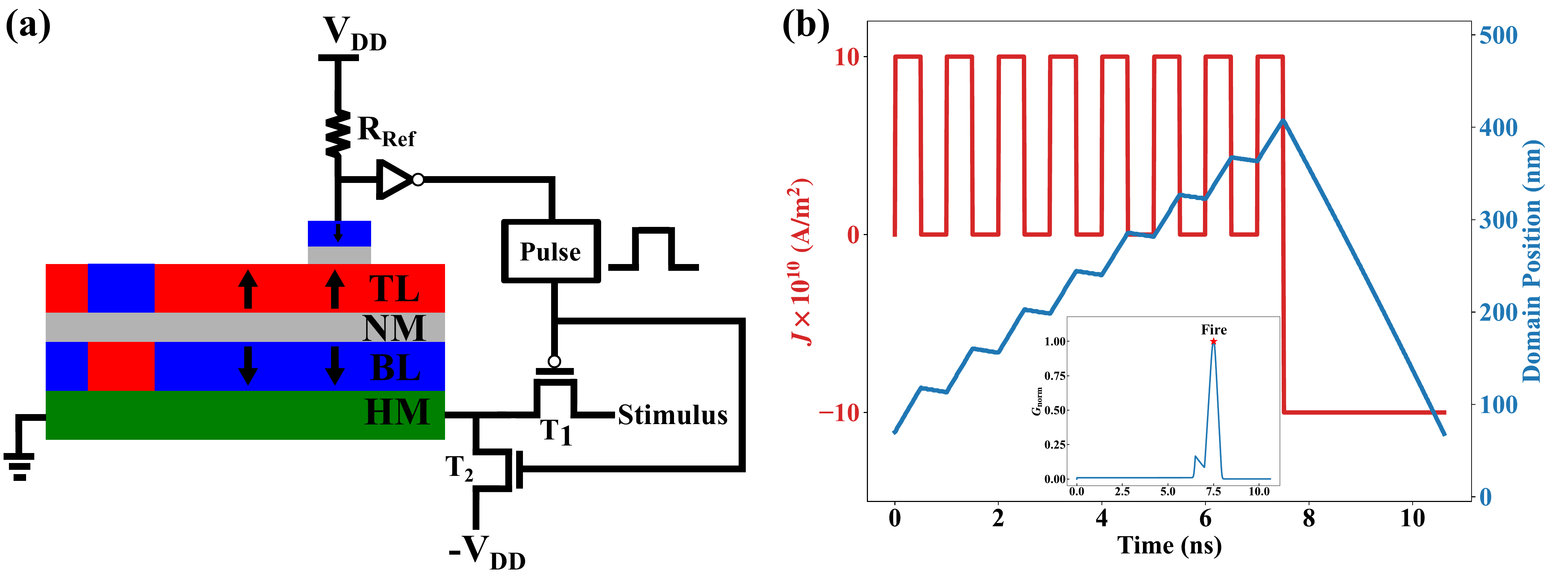}}
	\caption{(a) Schematics of the bilayer device with the HM layer attached to the BL, along with reset circuitry. (b) The plot of the input stimulus in the form of current pulse train that is injected to the HM, along with the reset current after the neuron fires (red), and the corresponding domain position (blue). Inset shows the time variation of normalized conductance in the detector region.}
	\label{Fig:Bilayer_HM_device_and_results}	
\end{figure}

The schematic of a neuron device that uses \ac{SOT} is shown in Fig.~\ref{Fig:Bilayer_HM_device_and_results}(a) where the input stimulus is injected via transistor $\mathrm{T_1}$, which is controlled by a pulse generator. 
Note that the device structure is similar to that in Fig.~\ref{Fig:Bilayer_device_wout_notch}(a) except for the \ac{HM} layer that is attached beneath the \ac{BL}.
The leaky behavior is achieved by the anisotropy gradient shown in Fig.~\ref{Fig:Bilayer_device_wout_notch}(b) just like in the previous neuron device.
Assume that the initial positions of the magnetic domains are similar to the devices described earlier.
When the magnetic domain is not underneath the detector region, the output of the pulse generator is at 0~V, which keeps $\mathrm{T_1}$ on and allows the input spikes to enter the device. 
Input spikes are applied as the square pulse train shown in Fig.~\ref{Fig:Bilayer_HM_device_and_results}(b) (red curve).
As \ac{SOT} devices are able to operate at lower current density \cite{khvalkovskiy2013matching, chureemart2021current}, we reduce the magnitude of the applied $J$ to $10\times10^{10}~\mathrm{A/m^2}$.
This charge current passes through the \ac{HM}, which generates a vertical spin current that exerts \ac{SOT} on the local magnetization in \ac{BL}.
The input stimulus moves the magnetic domain towards the detector region as shown by the plot of domain position versus time (blue graph) in Fig.~\ref{Fig:Bilayer_HM_device_and_results}(b).
The results confirm the integrate and leaky behavior of the neuron.
From the results, the magnetic domain requires 7.5~ns to reach the detector from rest and the energy consumed during the process is approximately 0.65~fJ.

Once the magnetic domain reaches the detector, the neuron fires.
The change in conductance in the detector region, $G_{\mathrm{norm}}$, versus time is shown in the inset of Fig.~\ref{Fig:Bilayer_HM_device_and_results}(b), where the red asterisk denotes the moment the neuron fires.
When this happens, the pulse generator is activated to output a 3.1~ns square pulse of magnitude $10\times10^{10}~\mathrm{A/m^2}$ as shown in Fig.~\ref{Fig:Bilayer_HM_device_and_results}(b).
This pulse deactivates $\mathrm{T_1}$ resulting in the neuron entering into the refractory period.
Simultaneously, the transistor $\mathrm{T_2}$ is activated and a reset current passes through the device.
This reset current returns the magnetic domains to their initial positions.
The energy consumed by this reset process is $\sim$0.5~fJ.
Thus, the total energy consumed by this device for the entire rest-to-fire-to-reset operation is 1.15~fJ.

\begin{figure}[!t]
	\centerline{\includegraphics[scale=0.4]{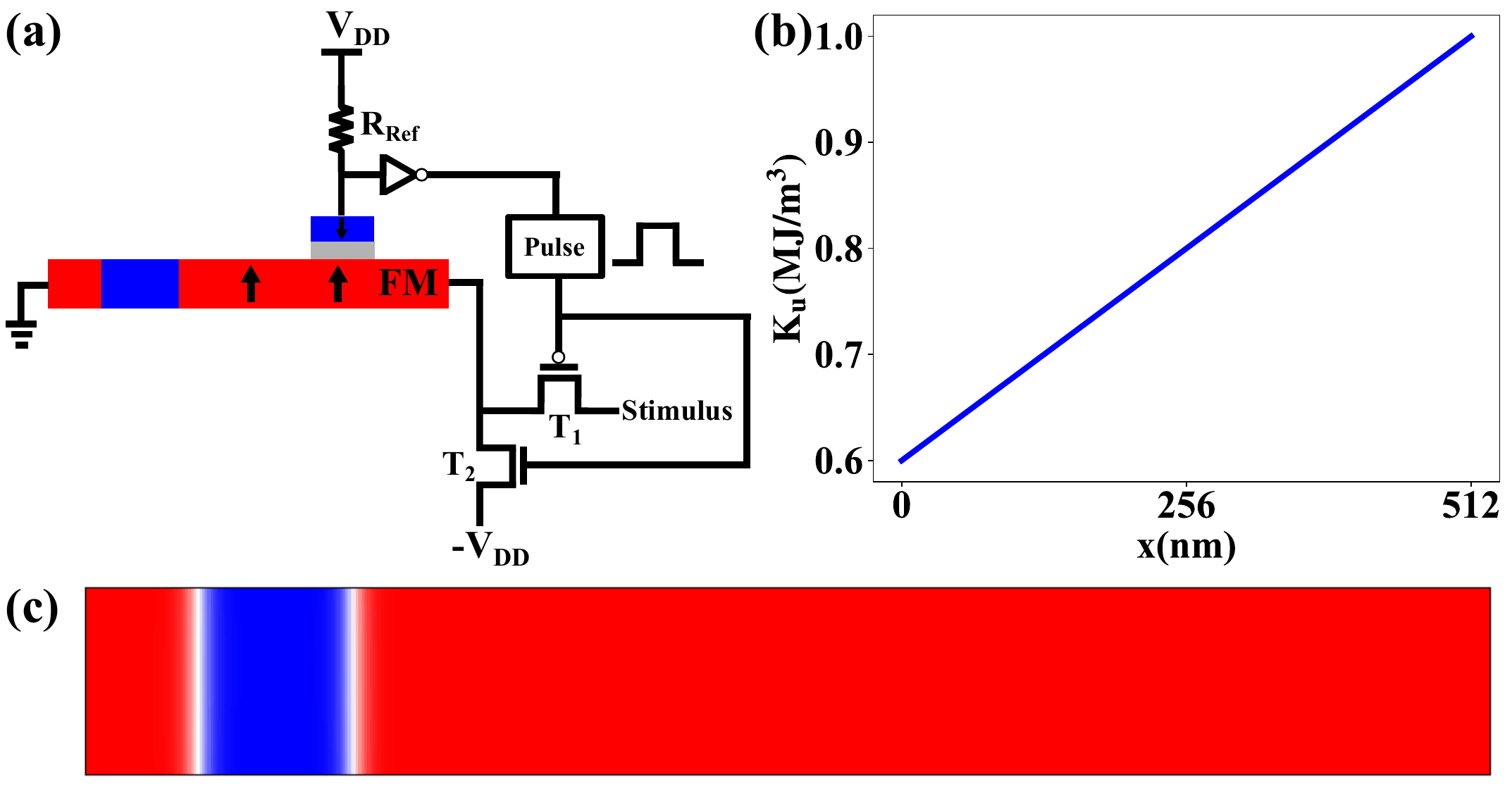}}
	\caption{(a) Schematics of the monolayer device along with reset circuitry. (b) $K_u$ profile along the length of the device. (c) Snapshot of the $m_z$ on the $x\mbox{-}y$ plane for the FM layer that is used as the initial magnetization profile for the monolayer neuron device.}
	\label{Fig:Monolayer_CIP_device}	
\end{figure}

Apart from the bilayer devices, the neuronal operation can also be achieved in the monolayer devices.
Consider the monolayer device shown in Fig.~\ref{Fig:Monolayer_CIP_device}(a). 
Similar to the previous devices, an anisotropy gradient along the length of the device shown in Fig.~\ref{Fig:Monolayer_CIP_device}(b) may be used to achieve the leaky behavior.
A \ac{MTJ} like detector is placed on the right side of the device.
A snapshot of $m_z$ on the $x\mbox{-}y$ plane is shown in Fig.~\ref{Fig:Monolayer_CIP_device}(c), showing the initial magnetization for the monolayer neuron device. 
\begin{figure}[!t]
	\centerline{\includegraphics[scale=0.19]{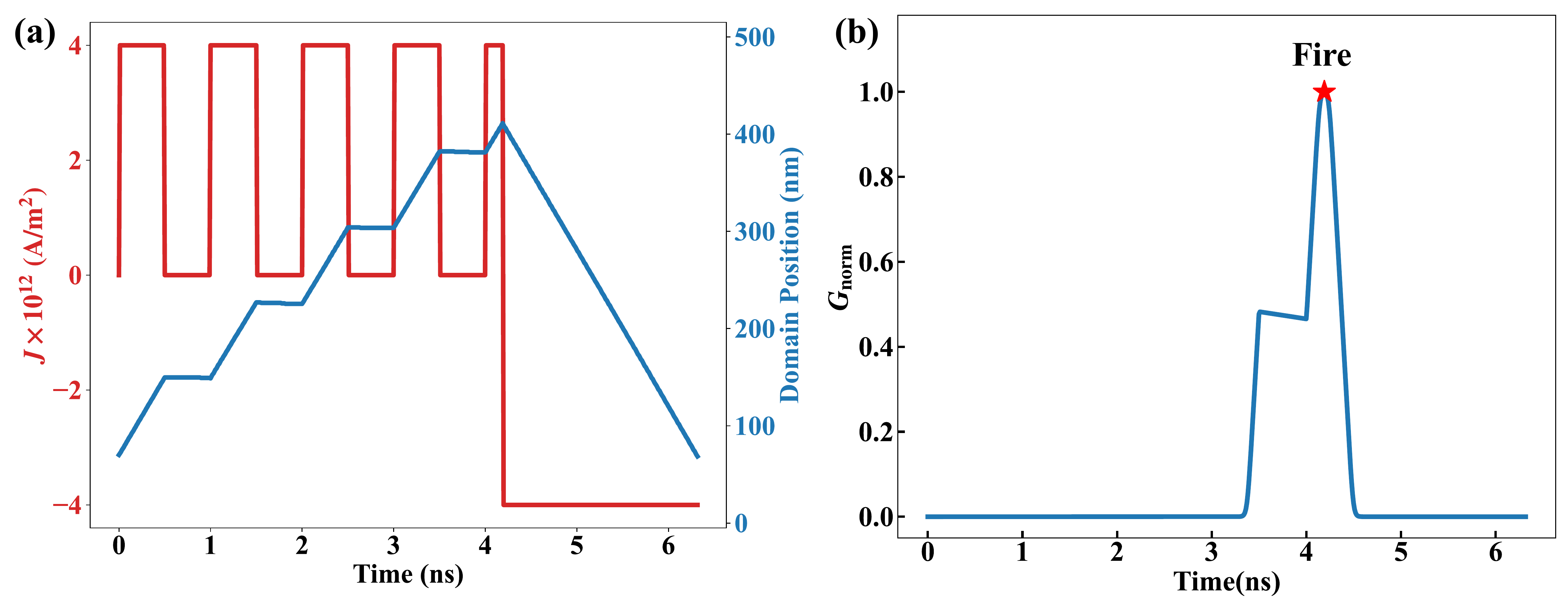}}
	\caption{(a) Plot of the current pulse train as the input stimulus injected to the FM layer, along with the reset current after the neuron fires, (b) Variation of the normalized conductance of the detector region with time. }
	\label{Fig_12}	
\end{figure}

To validate the neuronal operation of the device, a square pulse train of magnitude $4\times10^{12}~\mathrm{A/m^2}$, shown in Fig.~\ref{Fig_12}(a) is given as input stimulus to the device.
From the plot of the domain position versus time (blue curve), it may be observed that the magnetic domain moves toward and enters the detector after $\sim$4.19~ns.
The normalized conductance of the detector region versus time is shown in \ref{Fig_12}(b), where the firing event is marked by the red asterisk.
The energy consumed is $\sim$0.15~pJ.
When the magnetic domain reaches the detector, the neuron fires by generating a voltage spike at the voltage divider circuit.
This activates the pulse generator that generates a 2.12~ns square pulse of magnitude $4\times10^{12}~\mathrm{A/m^2}$ to deactivate transistor $\mathrm{T_1}$ and activate transistor $\mathrm{T_2}$.
Deactivating transistor $\mathrm{T_1}$ prevents input stimulus from affecting the device, which emulates the refractory period of the neuron.
Activating transistor $\mathrm{T_2}$ allows a reset current to reset the magnetic domain from the detector back to the initial position.
Our results show that 2.12~ns delay is needed to reset the neuron.
0.15~pJ of energy  is consumed during the reset process.
Thus, the leaky-integrate-and-fire operation of the device is validated and the total energy consumed for the entire rest-to-fire-to-reset operation for this monolayer device is about 0.3~pJ.

\begin{figure}[!b]
	\centerline{\includegraphics[scale=0.38]{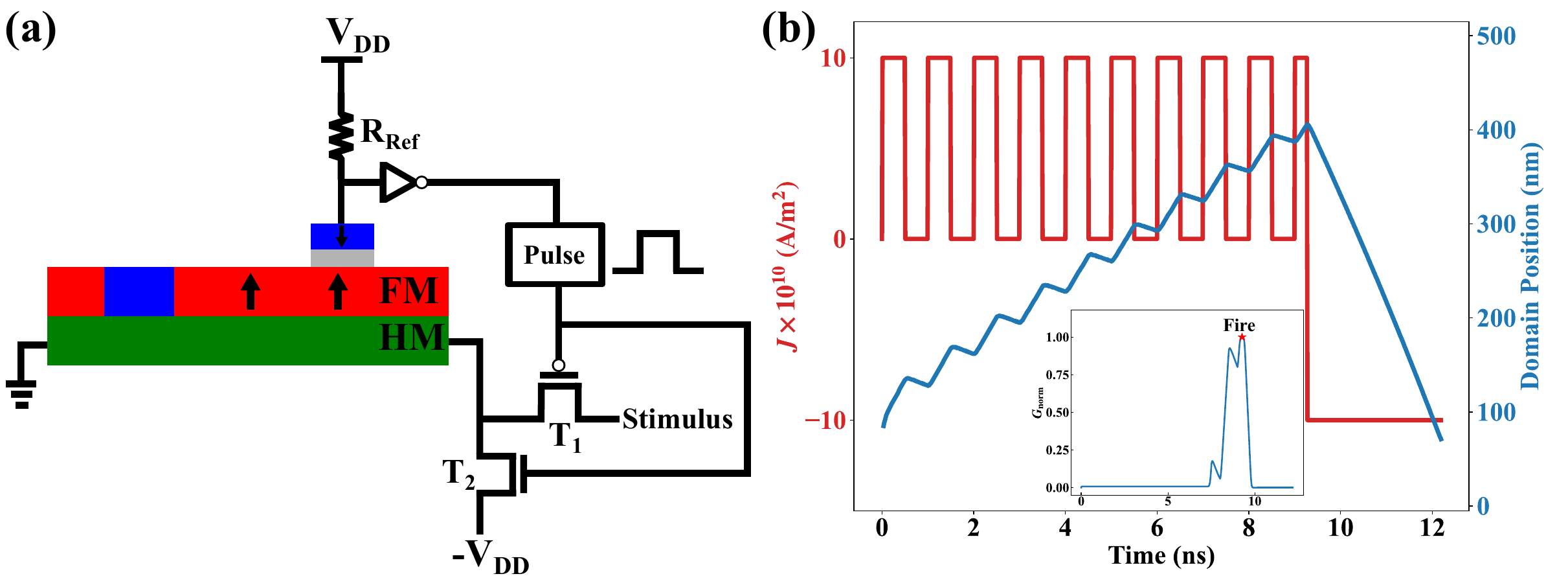}}
	\caption{(a) Schematic of the monolayer device where a \ac{HM} layer is attached to the \ac{FM}, along with the reset circuitry. (b) The plot of the current pulse train as the input stimulus is injected into the HM layer, along with the reset current after the neuron fires shown by the red plot, and variation of the magnetic domain position due to this current is shown in blue curve. Inset shows the time variation of normalized conductance in the detector region. }
	\label{Fig:Monolayer_CPP_device_results_inset}	
\end{figure}

Similarly, the energy consumed by the monolayer device may be reduced by attaching a \ac{HM} layer beneath the \ac{FM} layer.
The schematic of the device along with the reset circuitry is shown in Fig.~\ref{Fig:Monolayer_CPP_device_results_inset}(a). 
To validate the neuronal behavior of the device, consider the initial magnetization for this device to be as shown in Fig.~\ref{Fig:Monolayer_CIP_device}(c).
Input stimulus is applied to \ac{HM} in the form of a square pulse train of magnitude $10\times10^{10}~\mathrm{A/m^2}$, shown by the red curve in Fig.~\ref{Fig:Monolayer_CPP_device_results_inset}(b).
Our simulation results show that it takes 9.26~ns for the magnetic domain to move from its initial position to the detector region as shown by the blue curve in Fig.~\ref{Fig:Monolayer_CPP_device_results_inset}(b), which plots the position of the magnetic domain versus time.
The energy consumed during this process is $\sim$0.779 fJ.
The graph of normalized conductance of the detector region versus time is shown in Fig.~\ref{Fig:Monolayer_CPP_device_results_inset}(b).
Once the domain reaches the detector, the neuron fires and triggers the reset process which is similar to that for the previous device.
Our results indicate that the reset process takes 2.92~ns and consumes $\sim$0.47 fJ of energy.
Thus, the total energy consumption for the entire rest-to-fire-to-reset operation is $\sim$1.25~fJ.
\begin{table}[t]
	\centering
	\caption{\label{tab:table_summary}%
		Device simulation results
	}
	\label{Table:summary}
	\begin{tabular}{|c|c|c|c|}	
		\hline		
		\textrm{Device} & \textrm{Spin-torque} & \textrm{\makecell{Entire Operation\\  Time}} & \textrm{\makecell{Consumed\\ Energy}}\\
		\hline
		\makecell{Bilayer \\ with pinning site} & STT & 12 ns & 1.5 pJ\\
		\hline
		\makecell{Bilayer \\ without pinning site} & STT & 15 ns & 0.638 pJ\\
		\hline
		\makecell{Bilayer+HM \\ without pinning site} & SOT & 10.6 ns & 1.15 fJ\\
		\hline
		\makecell{Monolayer \\ without pinning site} & STT & 6.31 ns & 0.3 pJ\\
		\hline
		\makecell{Monolayer+HM \\ without pinning site}& SOT & 12.18 ns & 1.25 fJ\\
		\hline
	\end{tabular}	
\end{table}

A comparison of the spintronic neurons studied in this work is shown in Table~\ref{tab:table_summary}.
These results show that \ac{SOT} is a promising approach to reduce the total energy consumption of the spintronic neuron.
Moreover, the method of returning the magnetic domain from the detector to the initial position remains the most energy efficient approach due to relatively the large voltage and current needed to nucleate a new magnetic domain in the generator.
Furthermore, although the energy consumed for rest-to-fire operation can be in the sub-fJ regime, the energy consumed for the reset operation can be comparable.
This further motivates the need to design novel schemes that can further reduce the energy consumed by the rest process.

We believe that there remains a range of values for $J$ for the successful operation of these proposed devices, and it also affects the choice of reset timing. 
Obtaining the range needs a rigorous simulation for various values of $J$, which is beyond the scope of this work. 
Process variations are also worth investigating, as they provide an allowed range of parameter values for the successful operation of the proposed neuron devices. 
This also needs a rigorous simulation with the variation of different parameter values over a broad range.
In the current work, we intend to demonstrate various reset schemes for the neurons and compare the energy consumption between those schemes.
Thus, we omit the parameter variations in this current work and left as future work to pursue.

\section{Conclusions}\label{sec:conclusion}

In summary, several spintronic neuronal device concepts that have the ability to self-reset based on monolayer and \ac{SAF}-coupled bilayer systems were investigated.
Two methods of self-reset were studied to analyze their energy efficiency.
In one method, the magnetic domain representing the membrane potential of the neuron is nucleated at the reset position while the previous magnetic domain is annihilated.
In the second method, a reset current is utilized to return the magnetic domain to its initial position after the neuron fires.
Among schemes that were investigated, the devices utilizing \ac{SOT} consume lesser energy.
Furthermore, among the devices using \ac{SOT}, the bilayer device may achieve 8\% lower energy consumption than the monolayer device.
\bibliographystyle{IEEEtran}
\bibliography{References}

\end{document}